\documentclass[]{agujournal_arxiv}
\usepackage{wasysym} 

\journalname{Journal of Geophysical Research - Planets}

\begin{document}

%
%

\title{Constraining the magnitude of climate extremes from time-varying instellation on a circumbinary terrestrial planet}

%
%

 \authors{Jacob Haqq-Misra\affil{1}, Eric T. Wolf\affil{2}, 
William F. Welsh\affil{3}, Ravi Kumar Kopparapu\affil{4},
 Veselin Kostov\affil{4}, and Stephen R. Kane\affil{5}}

\affiliation{1}{Blue Marble Space Institute of Science, Seattle, Washington, USA.}
\affiliation{2}{University of Colorado Boulder, Boulder, Colorado, USA, USA.}
\affiliation{3}{San Diego State University, San Diego, California, USA.}
\affiliation{4}{NASA Goddard Space Flight Center, Greenbelt, Maryland, USA.}
\affiliation{5}{University of California Riverside, Riverside, California, USA.}


\correspondingauthor{Jacob Haqq-Misra}{jacob@bmsis.org}



%
%

\begin{abstract}
Planets that revolve around a binary pair of stars are known as circumbinary planets.
The orbital motion of the stars around their center of mass causes a periodic variation 
in the total instellation incident upon a circumbinary planet.
This study uses both an analytic and numerical energy balance model 
to calculate the extent to which
this effect can drive changes in surface temperature on circumbinary
terrestrial planets. We show that the amplitude of the temperature variation
is largely constrained by the effective heat capacity, which corresponds 
to the ocean-to-land ratio on the planet. Planets with large ocean fractions
should experience only modest warming and cooling of only a few degrees, 
which suggests that habitability cannot be precluded for such circumbinary planets.
Planets with large land fractions that experience extreme periodic forcing
could be prone to changes in temperature of tens of degrees or more, 
which could drive more extreme climate changes that inhibit continuously habitable conditions. 
\end{abstract}

%
%

\section{Introduction}
\label{sec:intro}

The recent discovery of giant planets in the habitable zone of close binary stars 
(\textit{e.g.}, \citet{welsh2017two} and references therein) has raised the possibility that
such systems could host terrestrial planets within the liquid water habitable zone. 
The circumstellar liquid water ``habitable zone'' has received broad attention for single-star 
systems (\textit{e.g.}, \citet{kasting1993habitable,abe2011habitable,kopparapu2013habitable,Kopparapu2014}, 
but terrestrial planets in binary systems may also be able to retain stable atmospheres
with habitable surface conditions. Previous studies have 
calculated the boundaries of the habitable zone in circumbinary systems (also known as 
P-type systems \citep{dvorak1984numerical}) 
by combining the spectral energy distributions of the host stars to determine
the orbital range that can maintain stable climates for an Earth-like planet 
\citep{kane2012habitable,haghighipour2013calculating,forgan2013assessing,
cuntz2013s,cuntz2015s,wang2019s,wang2019s2,georgakarakos2019enlargement}. 
Such studies demonstrate that 
circumbinary systems could host dynamically stable planets that maintain habitable
conditions, so long as the planet has a sufficiently dense atmosphere and a method
for recycling carbon between the atmosphere and interior (\textit{i.e.}, plate tectonics). 
Circumbinary planets have also been argued to have enhanced habitability prospects due to reduced stellar activity and XUV radiation \citep{Mason2013}. 

A circumbinary planet experiences changes in instellation 
as the binary pair orbit one another. The orbital separation of the binary pair
causes a periodic variation in incident radiation, which changes
both the intensity and distribution of
the incident spectral energy distribution \citep{forgan2015surface}.
In some cases more extreme variations in incident energy can occur when one star
eclipses another (as seen by the planet); however, while these eclipses are global in extent, they only have a duration of hours. 
The longer period variation in radiation from the binary pair is a time-dependent 
factor that could alter a planet's ability to sustain habitable conditions, even if it is otherwise
situated within the habitable zone. 
We refer to this periodic change in energy incident upon the planet as
the circumbinary ``gyration effect.''

Case studies of existing circumbinary systems suggest that this gyration effect may only 
exert a modest effect on global climate. (Such results corroborate other studies that have
generally found modest deviations from the mean flux approximation when considering 
variations in a planet's orbital eccentricity \citep{dressing2010habitable,bolmont2016habitability,way2017effects}.)
\citet{may2016examining} examine this periodic variation in instellation for Neptune-like circumbinary planets in the Kepler-47 system
by using both an energy balance model and an idealized general circulation model (GCM) to calculate the
maximum temperature changes expected. \citet{may2016examining} find that Neptune-like circumbinary 
planets would experience no more than one percent variation in temperature from the gyration effect.
\citet{popp2017climate} use a GCM to calculate climate variations for hypothetical
Earth-like planets in the Kepler-35 system and find that 
the global mean surface
temperature changes by a few degrees at most. These results suggest that the gyration effect is 
unlikely to preclude habitability for planets within the circumbinary habitable zone.
However, these previous efforts investigated planets with a uniform surface with a large
heat capacity, which is appropriate for an ocean-covered planet or Neptune analog but
may underestimate temperature variations on an Earth-like planet. 

In this study, we attempt to determine the extent to which the 
gyration effect is capable of driving significant changes in temperature
for dynamically-stable circumbinary planets that orbit within the habitable zone. 
We address cases where the planets transit main-sequence stars, 
\textit{i.e.}, cases similar to the Kepler circumbinary planets.
We use an analytic energy balance model to show that the effective heat capacity
of a planet exerts the greatest control on the magnitude of the temperature variation
that results from the circumbinary gyration effect.
We then use a numerical energy balance model to show that the amplitude of 
time-dependent changes in temperature is greatest for planets with a large land
fraction (and thus a lower effective heat capacity).
Our results suggest that circumbinary planets with a lower ocean-to-land surface fraction 
that undergo strong periodic forcing are more likely to experience extreme climate change.

\section{Maximum variation from periodic forcing}

As the distances between a circumbinary planet and its host stars continuously vary, the planet experiences
a change in instellation throughout its orbit. 
The amplitude of the change depends on the orbits of the binary stars and the planets 
(specifically the luminosity of each star, semi-major axes, and eccentricities; throughout
this paper we assume co-planar orbits). The timescale of the change depends on the orbital periods of the binary, the planet, and long-term precession of the planet's orbit.
Of the known Kepler circumbinary planets, the shortest period binary is that of Kepler-47 with a period of 7.4 days, 
with planets b, c, and d on 49.5, 303, and 187 day orbits \citep{orosz2019discovery}.
The longest period host binary is that of Kepler-16, with an orbital period of 41 days \citep{Doyle2011} and a planet 
on a $\sim$ 229 day orbit.
Note that the longer the period of the binary, 
the lower the probability of transit, and so
the more difficult it is to detect a circumbinary planet.
Current observations thus do not provide much constraint on the maximum orbital period of the binary. 
However, the stability criteria of \citet{holman1999long} requires that 
the binary must have a period shorter than approximately one third of the planet's orbital period 
in order for the planet to be dynamical stable (assuming a circular planetary orbit). 
In this study, we apply this stability criterion between the binary orbital period, $P_{bin}$, and
the orbital period of the circumbinary planet, $P_{cbp}$,
in order to constrain the extent of temperature variation on Earth-like circumbinary planets.

The maximum variation in instellation spans a wide range 
across the observed Kepler circumbinary systems. 
The most variable is Kepler-34b, which is not in the habitable zone (too hot).
Kepler-34b is the most eccentric of known transiting circumbinary planets (\citet{welsh2012}), 
and its eccentric orbit ($e_{cbp}=0.182)$ is a significant factor driving changes in instellation. 
Kepler-35b, also not in the habitable zone, has a much lower eccentricity 
than Kepler-34b ($e_{cbp}=0.042$), which leads to a semiamplitude 
of about 15\%. 
The circumbinary systems that are in the habitable zone 
(Kepler-16, 47, 453, 1647) have similar instellation semiamplitudes, less than
20\% on a timescale of a few planetary orbits.
However, on a longer timescale ($\sim$ decades to centuries) 
planet orbit precession can play a significant role:
Kepler-16b has the most circular orbit ($e_{cbp}=0.0069$) of the known circumbinary planets, 
but experiences a range in semiamplitude variation from about 17\% to 28\% due to evolution 
of its orbit. This suggests that a semiamplitude of 30\% is certainly reasonable for a 
nearly circular orbit, and 
higher values as the eccentricity increases. 
The most eccentric Kepler circumbinary planet is the non-transiting planet KIC 7821010 b 
with $e \sim 0.35$ [\textit{Priv. Comm.}]. Similar to that of Kepler-34 b, its change in instellation is $\sim$90\%. 
But unlike Kepler-34 b, KIC 7821010 b orbits within 
the habitable zone (most of the time) with an average instellation equal to $\sim$ 68\% of the solar flux.

\begin{figure}[ht!]
\centerline{\includegraphics[width=5.5in]{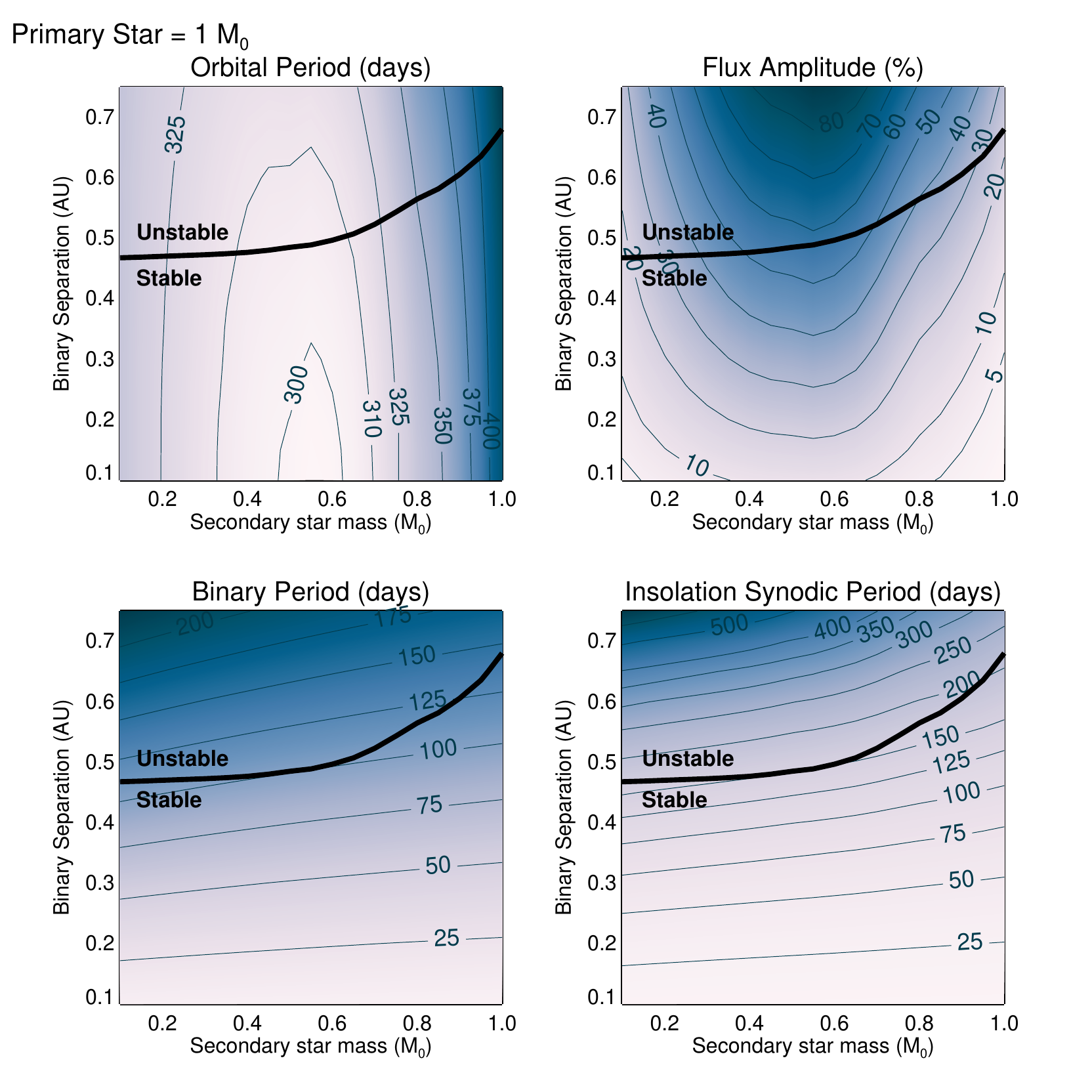}}
\caption{The orbital period of the planet (top left), flux amplitude (top right), 
binary orbital period (bottom left), and insolation synodic period 
at the planet's orbit (bottom right) depend on the binary separation
and the mass of the secondary star. Calculations assume a $1\,M_{\astrosun}$ primary
and a planet in a circular circumbinary orbit with a time mean
instellation of 1360\,W\,m$^{-2}$.
The dark curve indicates the maximum binary separation that permits a stable
orbitng planet. The maximum flux amplitude of 55\% occurs for a 
$0.5\,M_{\astrosun}$ secondary star, which corresponds to 
a planetary orbital period of about 305 days, 
a binary orbital period of about 100 days, and
an insolation synodic period of about 150 days on the planet.
\label{fig:orbits}}
\end{figure}

We first estimate the maximum amplitude of the gyration effect 
for a hypothetical circumbinary system with a planet that receives the same
time-mean value of stellar insolation as Earth today. 
We keep the mass of the primary star fixed at $1\,M_{\astrosun}$ and vary
the mass of the secondary from 0.1 to $1\,M_{\astrosun}$, with
a binary separation ranging from 0.1 to 0.75\,AU. We simplify our calculations
by assuming a non-eccentric orbit and solve Kepler's laws
using stellar mass-luminosity functions \citep{pecaut2013intrinsic,pecaut2012revised}.
The top left panel 
of Fig. \ref{fig:orbits} shows 
the orbital period of a planet in such a circumbinary
system (following Kepler's laws), where the incident stellar flux 
on the planet has been fixed at the present-day Earth value of 1360\,W\,m$^{-2}$. 
The black curve indicates the maximum binary separation that permits a planet
to remain in a stable orbit at this incident flux, following the
$P_{bin} \approx P_{cbp} / 3$ stability criteria \citep{holman1999long}.
The amplitude of flux variation from the circumbinary pair
is shown in the top left panel of Fig. \ref{fig:orbits}, 
which indicates a maximum flux amplitude of about 55\%
with a secondary mass of about 0.5\,M$_{\astrosun}$ and 
a binary separation of about 0.5\,AU. 

We next determine the period of variation in instellation that corresponds to 
this hypothetical circumbinary system with a 1\,M$_{\astrosun}$ primary
and 0.5\,M$_{\astrosun}$ secondary.
The maximum flux amplitude occurs
with a binary orbital period of about 100\,days, as shown in the bottom
left panel of Fig. \ref{fig:orbits}. The planet in such a system
has an orbital period of about 305\,days, so the period of variation in flux 
is determined by the combined motion of the binary pair and circumbinary planet.
The resulting insolation synodic period is shown in the bottom right panel 
of Fig. \ref{fig:orbits}, which indicates a period of about 150 days. 
In our model calculations that follow, we will consider the response
of a terrestrial planet in such a circumbinary system, 
with the period of the circumbinary gyration effect at 150 days 
and the maximum amplitude of flux variation at 50\%.
We specifically choose this extreme instellation variation scenario 
to examine the maximum response of the circumbinary planet's climate, 
and to answer the question, ``Are climate variations due to the gyration effect 
negligible or significant?'' But note that for developing our intuition, via 
the analytic energy balance calculations in the next section, we use circular orbits. 
Much more extreme instellation variations are possible for highly eccentric orbits.

\section{Analytic energy balance model}\label{sec:earthlike}

This section examines the amplitude of temperature variation due to 
the time-dependent circumbinary gyration effect. 
The analytic energy balance calculations 
shown below demonstrate that the effect typically causes temperature 
to oscillate by only a few degrees 
for most cases, which is consistent with previous
results \citep{may2016examining,popp2017climate}. At the same time,
planets with a much lower effective heat capacity may experience tens of
degrees of temperature change.

The rate of change of surface temperature, $dT/dt$, on a terrestrial
planet depends upon the incoming stellar energy, $S$, the planetary Bond albedo,
$\alpha$, and the outgoing infrared radiative flux, $F_{IR}$. We
can express this relationship as
\begin{equation}
C\frac{dT}{dt}=S\left(1-\alpha\right)-F_{IR},\label{eq:EBM1}
\end{equation}
where $C$ is effective heat capacity of the surface and atmosphere. 
A simple representation of $F_{IR}$ is the linear
parameterization $F_{IR}=A+BT$, where $A$ and
$B$ are infrared flux constants (with $T$ in Celsius).
For this problem, we are interested in a periodic variation
of the incoming stellar energy. This serves as an analogy to the changes
in flux for a circumbinary planet when its host stars 
orbit 
one another. 
We choose a simple periodic function:
\begin{equation}
S=S_{0}\left(1 + \kappa \sin\omega t\right),\label{eq:stellarflux}
\end{equation}
where $S_{0}$ is a constant value of incident stellar flux, 
$\omega$ is the angular frequency of flux variation (with $\omega=2\pi/P_{bin}$), and
$\kappa$ is the amplitude of flux variation. 
This function is constructed to begin with a flux of $S=S_{0}$ at $t=0$ and 
recover a single-star solution when $\kappa=0$.
We acknowledge that this sinusoidal
representation of stellar flux implies a circular planetary orbit ($e_{cbp}=0$), which would be dynamically
unstable in an actual circumbinary system; however, it is possible for planet orbits
to remain stable with small non-zero eccentricities.
(One way to extend this analytic model would be 
to develop the forcing as a real-valued Fourier series \citep{popp2017climate},
with a solution that corresponds to the sum of several modes. 
The single-mode solution developed in this paper indicates the qualitative behavior
expended from more realistic, multi-modal circumbinary orbits.)
We will proceed by emphasizing
that these calculations are intended to place theoretical limits on the impact of time-varying
instellation on climate by using an idealized representation of the circumbinary flux variation.
We can then write Eq. (\ref{eq:EBM1}) as 
\begin{equation}
C\frac{dT}{dt}=S_{0}\left(1-\alpha\right)\left[1+\kappa\sin\left(\omega t\right)\right]-\left(A+BT\right).\label{eq:EBM2}
\end{equation}
We can solve this ordinary differential equation in order to obtain
an expression for $T\left(t\right)$ that depends upon $C$, $\omega$,
and $\kappa$. We assume that $\alpha$ is constant,
analogous to an ice-free and ocean-covered planet (i.e., no ice-albedo feedback).

The analytic solution to Eq. (\ref{eq:EBM2}) is given in the Appendix, which 
yields an expression for the steady-state solution:
\begin{equation}
T\left(t\right)\approx\frac{S_{0}\left(1-\alpha\right)\kappa}{\omega C}\cos\left(\omega t\right)+T_{0}.\label{eq:steadystate}
\end{equation}
The amplitude of Eq. (\ref{eq:steadystate}) represents the magnitude of temperature perturbations
from circumbinary forcing. It is noteworthy that the infrared flux constants $A$ and $B$ are
absent from Eq. (\ref{eq:steadystate}), with the effective heat capacity as the primary
planetary property that affects the amplitude of the gyration effect. Although 
the mean temperature $T_0$ is determined by properties of the atmosphere's composition
(through the greenhouse effect, as described by $A$ and $B$), the amplitude of circumbinary
change in this analytic model is insensitive to changes in the greenhouse effect.

In order to apply Eq. (\ref{eq:steadystate}) to the scenario of a
circumbinary planet in the habitable zone, we assume Earth-like conditions
of $T_{0}=15\,{}^{\circ}\text{C}$ and $C=2.1\times10^{8}\,\text{J m}^{-2}\,^{\circ}\text{C}^{-1}$.
We are also interested in small perturbations in temperature that result
from variations in instellation as the binary pair rotates, so we
ignore any changes in albedo due to ice growth and assume a constant
value of $\alpha$. We obtain a value for albedo by setting
Eq. (\ref{eq:EBM2}) to a steady state ($dT/dt=0)$ with a fixed single
star ($\omega=0$) and solving for $\alpha$ when $T=T_{0}$, which
gives $\alpha=1-\left(A+BT_{0}\right)/S_{0}\approx0.31$. 
(This expression for albedo assumes Earth-like values for the thermal emitted flux parameters, 
$A=203\text{ W m}^{-2}$ and $B=2\text{ W m}^{-2}\text{ }^{\circ}\text{C}^{-1}$, 
which are based on northern hemisphere observations \citep{north1981energy}.)
We use this
value of $\alpha$ to evaluate Eq. (\ref{eq:steadystate}).
Given the assumptions that went into this calculation, it should not be surprising that this is
in fact similar to the measured value of $\alpha \approx 0.29$ for Earth (\textit{e.g.} \citet{Stephens2015}).

\begin{figure}[ht!]
\centerline{\includegraphics[width=5.5in]{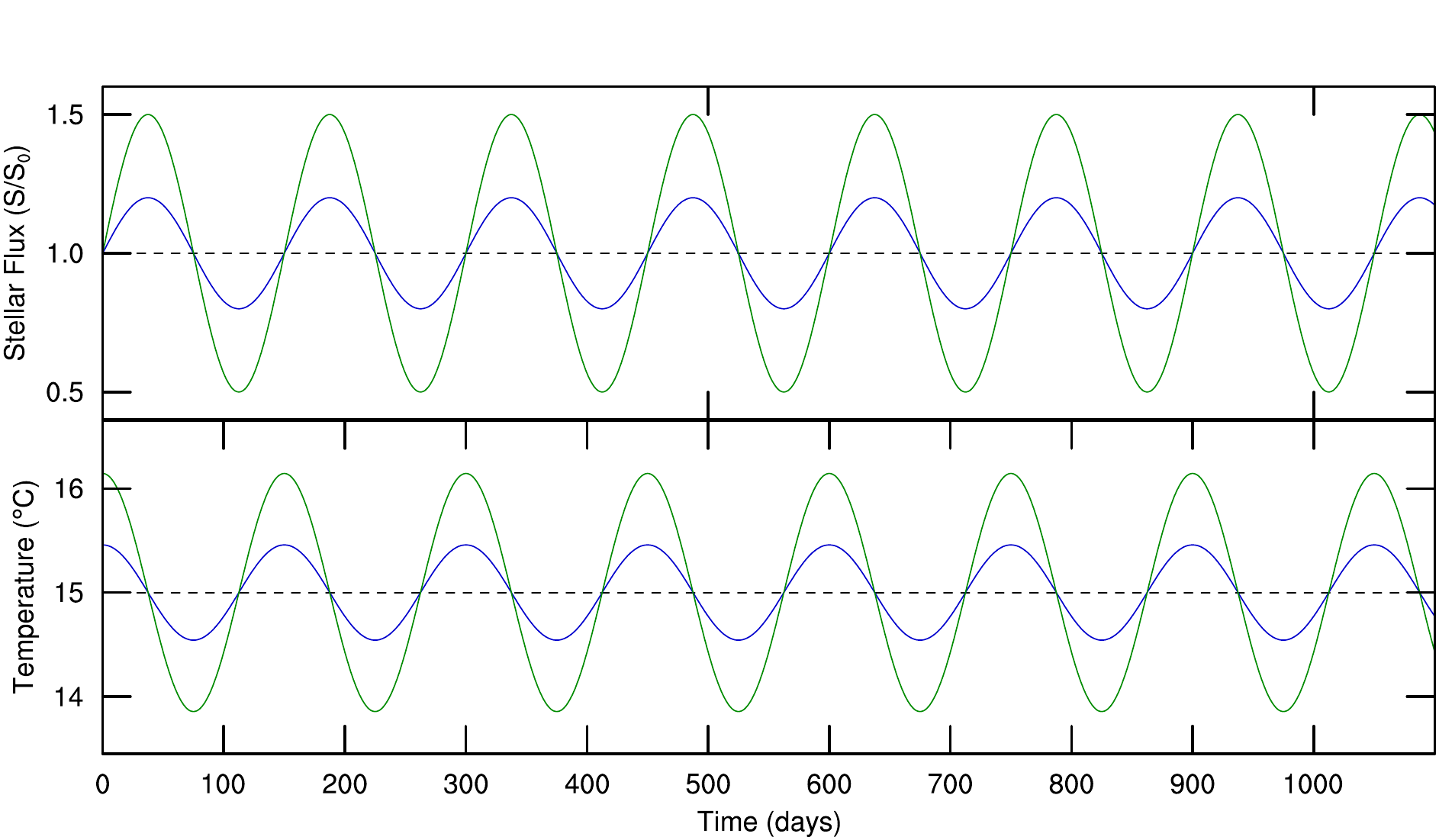}}
\caption{Steady state variations with time using the analytic EBM 
of the relative stellar flux (top) and
mean planet temperature (bottom) for a fictitious 
1\,M$_{\astrosun}$ primary and 0.5\,M$_{\astrosun}$ secondary case with $P_{bin}=150$ days.
Colored curves show a single star
control case with $\kappa=0$ (dashed black) and circumbinary
cases with $\kappa=0.2$ (blue) and $\kappa=0.5$ (green).
\label{fig:analytic2}}
\end{figure}

Examples of solutions given by Eq. (\ref{eq:steadystate}) are shown in Fig. \ref{fig:analytic2},
which includes the time variation in stellar flux (top panel) and steady-state
temperature (bottom panel) for several cycles. These cases
assume $P_{bin}=150$ days ($\omega\approx0.04\,\text{rad day}^{-1}$),
with a single star control case ($\kappa=0$) and two circumbinary
cases with small amplitude ($\kappa=0.2)$ and extreme amplitude ($\kappa=0.5$).
The variation in mean planet temperature shows a lag relative to the variations in stellar flux. 
The lag timescale, $t_{lag}$, is the difference in the time 
between the maxima of the stellar forcing function from Eq. (\ref{eq:stellarflux})
and temperature from Eq. (\ref{eq:fullgeneral}), which we can 
express as
\begin{equation}
t_{lag}=\frac{1}{\omega}\left(\frac{\pi}{2}+\arctan\frac{B}{\omega C}\right)\approx\frac{\pi}{2\omega}=\frac{P_{bin}}{4},
\end{equation}
where we have made the simplifying assumption $\omega{C}\gg{B}$.
The lag time for climate thus depends only upon the circumbinary orbital period, rather than 
properties of the planet's radiative transfer variable described by $B$.
At larger circumbinary orbital periods, the lag timescale increases and enables
a delayed response of climate to periodic changes in instellation. The $P_{bin}=150$ day circumbinary
period considered in Fig. \ref{fig:analytic2} corresponds to a lag timescale of $t_{lag}=38$ days, which
is long enough to induce a periodic temperature change of one to several degrees.

These calculations indicate that
the gyration effect of periodic variation about the mean in
incident stellar flux is capable of driving periodic change in planetary temperature.
The angular frequency of flux variation, $\omega$, (or equivalently, the period $P_{bin}$) determines
the lag between flux variation and temperature response, which can substantially alter
the magnitude of variation in $T$. In addition,
the two parameters that influence temperature in this model are the
amplitude of circumbinary forcing, $\kappa$, and the effective
heat capacity, $C$. Physically, a larger value of $\kappa$
corresponds to a larger flux variation from the binary pair. A larger
value of $C$ corresponds to a planet with greater average heat capacity
that is less sensitive to time-varying
changes in energy (such as a planet with a large fraction of ocean coverage).

\begin{figure}[ht!]
\centerline{\includegraphics[width=4.0in]{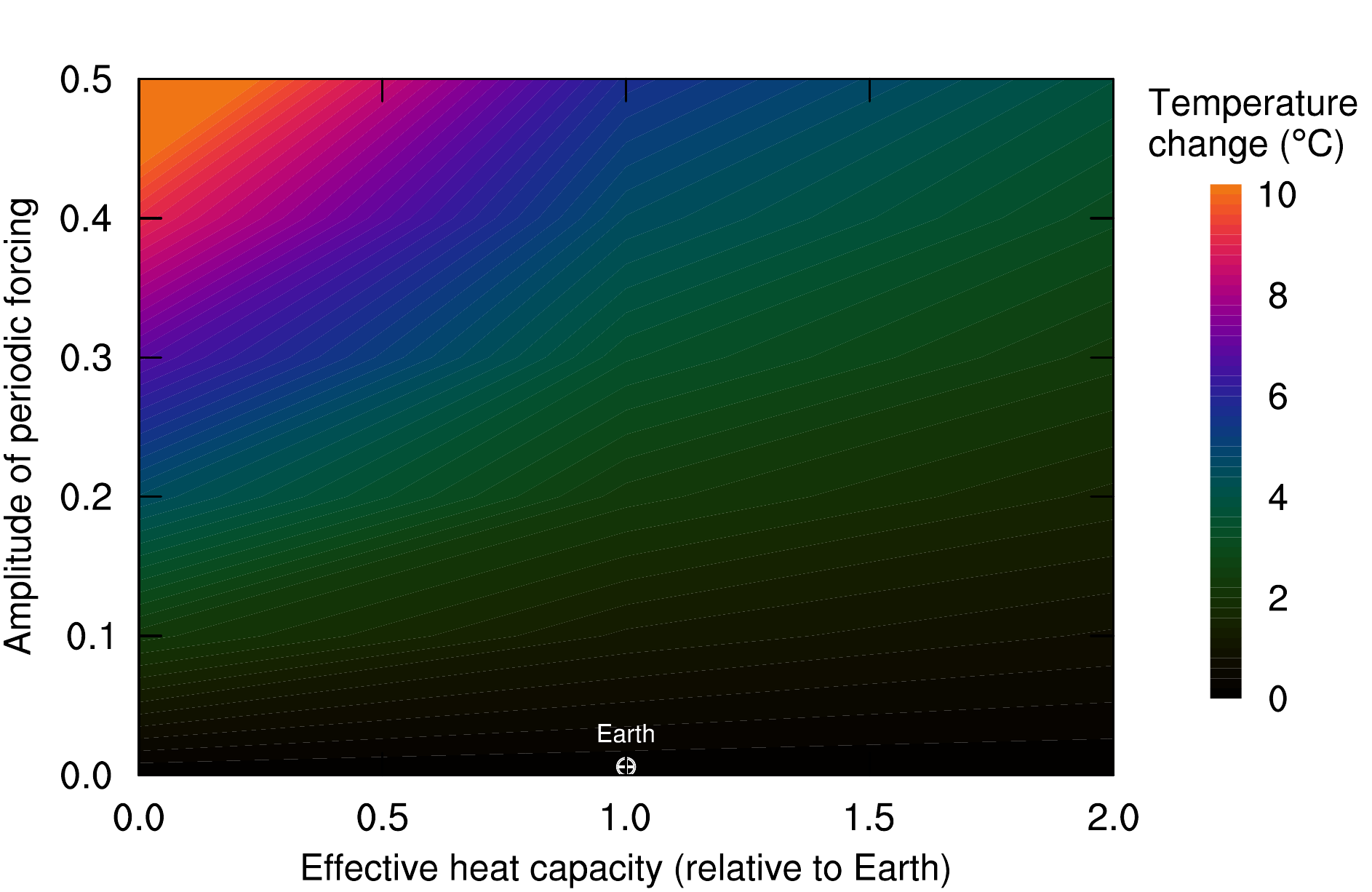}}
\caption{Maximum temperature change using the analytic EBM 
after reaching a steady state, shown over a parameter space of
effective heat capacity and the amplitude of periodic forcing.
The white circle indicates the heat capacity and constant forcing of present-day Earth.
\label{fig:contour}}
\end{figure}

The maximum temperature variation that can reasonably
be attained through this model is given by the amplitude of the cosine term 
in Eq. (\ref{eq:steadystate}). 
This maximum amplitude is shown in Fig. \ref{fig:contour}
as a ``heat map'' plot 
over the parameter space of $C$ and $\kappa$, with $P_{bin} = 150$ days,
The largest changes in temperature occur for planets with low effective
heat capacity and a large amplitude of periodic
forcing. Most of the parameter space shows only a few 
degrees total warming (and, analogously, the same amount of cooling), which
corroborate previous studies \citep{may2016examining,popp2017climate}. 
This suggests that the gyration effect
is unlikely to substantially undermine the habitability of an Earth-like
planet under the most typical conditions. But the upper-left quadrant of 
Fig. \ref{fig:contour} also indicates
that tens of degrees of variation could be possible for atmospheres with low effective
heat capacity that experience large variations in stellar forcing. Under
extreme conditions, circumbinary planets are capable of experiencing dramatic
shifts in temperature. 

This simple analytic model shows that a planet's susceptibility to 
the circumbinary gyration effect is primarily controlled by its effective heat capacity.
In the case of terrestrial planets, effective heat capacity is lowest for a dry land planet
and increases for planets with greater ocean coverage and depth.
The coefficients of the outgoing infrared radiative flux, $A$ and $B$, can be altered
to describe atmospheres of various compositions, based upon the net greenhouse warming, but
these parameters do not affect the steady-state gyration effect described by Eq. (\ref{eq:steadystate}).
The gyration effect may therefore affect planetary habitability under
certain scenarios. Circumbinary planets similar to Earth in terms of effective heat capacity 
(i.e., a similar ocean to land distribution), with low to moderate variation in instellation,
may not experience significant variations in surface temperature. The gyration effect cannot
preclude habitability on such planets. Conversely, terrestrial planets with lower effective 
heat capacity (i.e., a lower ocean to land ratio) under larger instellation variation
may find their global climates dominated by the changing flux from the binary.
These extreme circumbinary systems may find themselves driven into climate regimes
that diminish habitable conditions.

\section{Numerical energy balance model}
The analytic EBM used in the previous section demonstrates the significance of effective heat 
capacity as the primary planetary property that determines the response to
circumbinary forcing. The analytic model assumes a fixed value of albedo, which thereby
negates any ice-albedo feedback that could occur on such a planet; however
planets that experience
a circumbinary gyration effect with magnitude of a few degrees or greater could experience
periodic changes in ice coverage that alter albedo and thus affect the total energy balance. 
The analytic model also constrains the planet to a single point, which neglects the contribution
of meridional energy transport from equator to poles. 
We therefore continue our examination of the circumbinary gyration effect using a numerical
EBM that accounts for ice-albedo feedback, meridional energy transport, and the effect of topography on 
effective heat capacity.

The numerical EBM \citep{haqq2014damping} is modified to include the circumbinary forcing
term from Eq. (\ref{eq:stellarflux}). The energy balance equation for this model can 
then be written in terms of latitude, $\theta$, as:
\begin{equation}
C\frac{\partial T}{\partial t}=S_{0}\left(1-\alpha\right)\left[1+\kappa\sin\left(\omega t\right)\right]-\left(A+BT\right)+\frac{1}{\cos\theta}\frac{\partial}{\partial\theta}\left(D\cos\theta\frac{\partial T}{\partial\theta}\right).\label{eq:steadystatenumerical}
\end{equation}
The last term in Eq. (\ref{eq:steadystatenumerical}) accounts 
for the efficiency of meridional energy transport by using $D$ as a diffusive parameter, 
with $D=0.38$\,W m$^{-2}$ K$^{-1}$ in this study as a fixed value applicable to 
present-day Earth conditions. Albedo is defined as a function of temperature so that 
$\alpha = 0.3$ for unfrozen land or ocean ($T \ge 263.15$ K) and 
$\alpha = 0.663$ for any frozen surface ($T \le 263.15$ K).
(Note that this threshold is ten degrees below freezing to indicate the formation
of permanent surface ice.)
The thermal emitted flux parameters are set to $A=203\text{ W m}^{-2}$ and 
$B=2\text{ W m}^{-2}\text{ }^{\circ}\text{C}^{-1}$.
The EBM decomposes the planet into 18 equally-spaced latitudinal zones, assumes a starting
surface temperature profile, and then numerically solves Eq. (\ref{eq:steadystatenumerical}) 
by using a time step of $\Delta t = 8.64\times10^3$ s = 1\,day. 
This model assumes a circular orbit for purposes of comparing directly with the analytic solution, 
while noting that planets in circumbinary systems will necessarily have non-zero eccentricity. 
We initially keep planetary obliquity fixed at zero degrees in order to eliminate any seasonal 
effects, although we later consider the effect of seasons (following \citet{gaidos2004seasonality})
on circumbinary planets with Earth-like axial tilt.

Effective heat capacity, $C$, is defined in the numerical EBM
as a function of both latitude and temperature, 
which represents the contributions from unfrozen land, unfrozen ocean, and ice-covered surface.
Letting $f_o$ and $f_i$ represent the respective fraction of ocean and ice
at each latitudinal zone, we can write the zonally averaged heat capacity as
\begin{equation}
C = \left(1-f_i\right)C_l + f_o \left[\left(1-f_i\right)C_o + f_i C_i \right],\label{eq:heatcapacity}
\end{equation}
where $C_l = 5.25\times10^6$\,J\,m$^{-2}$\,K$^{-1}$ is the heat capacity of continental land,
$C_o = 40\,C_l = 2.1\times10^8$\,J\,m$^{-2}$\,K$^{-1}$ is the heat capacity over a wind-mixed 50\,m
ocean layer, and $C_i = 2\,C_l = 1.05\times10^7$\,J\,m$^{-2}$\,K$^{-1}$ is the heat capacity
over ice \citep{fairen2012reduced}. Eq. (\ref{eq:heatcapacity}) enables the numerical EBM
to vary effective heat capacity according to surface conditions on the planet as a function
of latitude.

\begin{figure}[ht!]
\centerline{\includegraphics[width=5.5in]{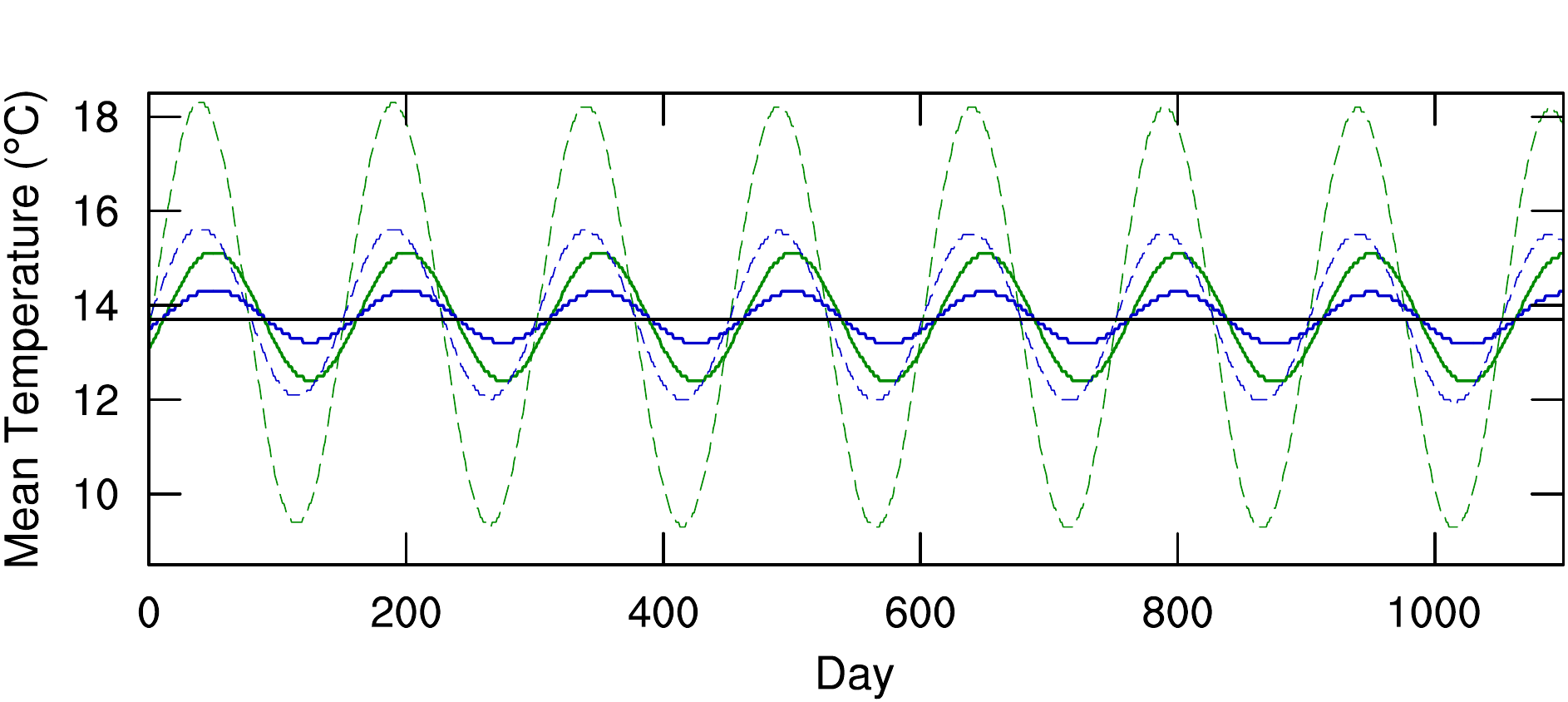}}
\caption{Steady state variations of mean planet temperature with time using the numerical EBM 
for a fictitious 
1\,M$_{\astrosun}$ primary and 0.5\,M$_{\astrosun}$ secondary case with $P_{bin}=150$ days.
Colored curves show a single star
control case with $\kappa=0$ (dashed black) and circumbinary
cases with $\kappa=0.2$ (blue) and $\kappa=0.5$ (green), all with zero obliquity.
Solid curves indicate global aquaplanet conditions with no topography, while dashed curves 
indicate an equatorial supercontinent as topography.
\label{fig:dailyavgtemp}}
\end{figure}

\subsection{Periodic temperature variations}

Steady-state solutions of Eq. (\ref{eq:steadystatenumerical}) 
are shown as mean temperature in Fig. \ref{fig:dailyavgtemp}
for the same $P_{bin} = 150$ day circumbinary system considered with the analytical model. 
The single star control case is shown as a black line on Fig. \ref{fig:dailyavgtemp}, along with
circumbinary cases with moderate forcing ($\kappa = 0.2$, blue curves) and 
strong forcing ($\kappa = 0.5$, green curves). Solid curves correspond to global aquaplanet conditions
with no land ($f_o = 1.0$), while dashed curves include topography as an equatorial supercontinent 
with ocean at the poles ($f_o = 0.7$). The aquaplanet cases plotted in Fig. \ref{fig:dailyavgtemp}
are comparable with the analytic solutions shown in Fig. \ref{fig:analytic2}, which show
a temperature oscillation of about a degree for $\kappa = 0.2$ and 
about three degrees for $\kappa = 0.5$. 
The amplitudes of these aquaplanet solutions with the numerical EBM 
are slightly greater than the analytic EBM as the result of enhanced climate
sensitivity from the inclusion of ice-albedo feedback and meridional energy transport.
The equatorial supercontinent case shows a much larger amplitude
of variation of about four degrees for $\kappa = 0.2$ and 
nearly ten degrees for $\kappa = 0.5$, due to the reduction in effective
heat capacity along the planet's equatorial belt. 

\begin{figure}[ht!]
\centerline{\includegraphics[width=5.5in]{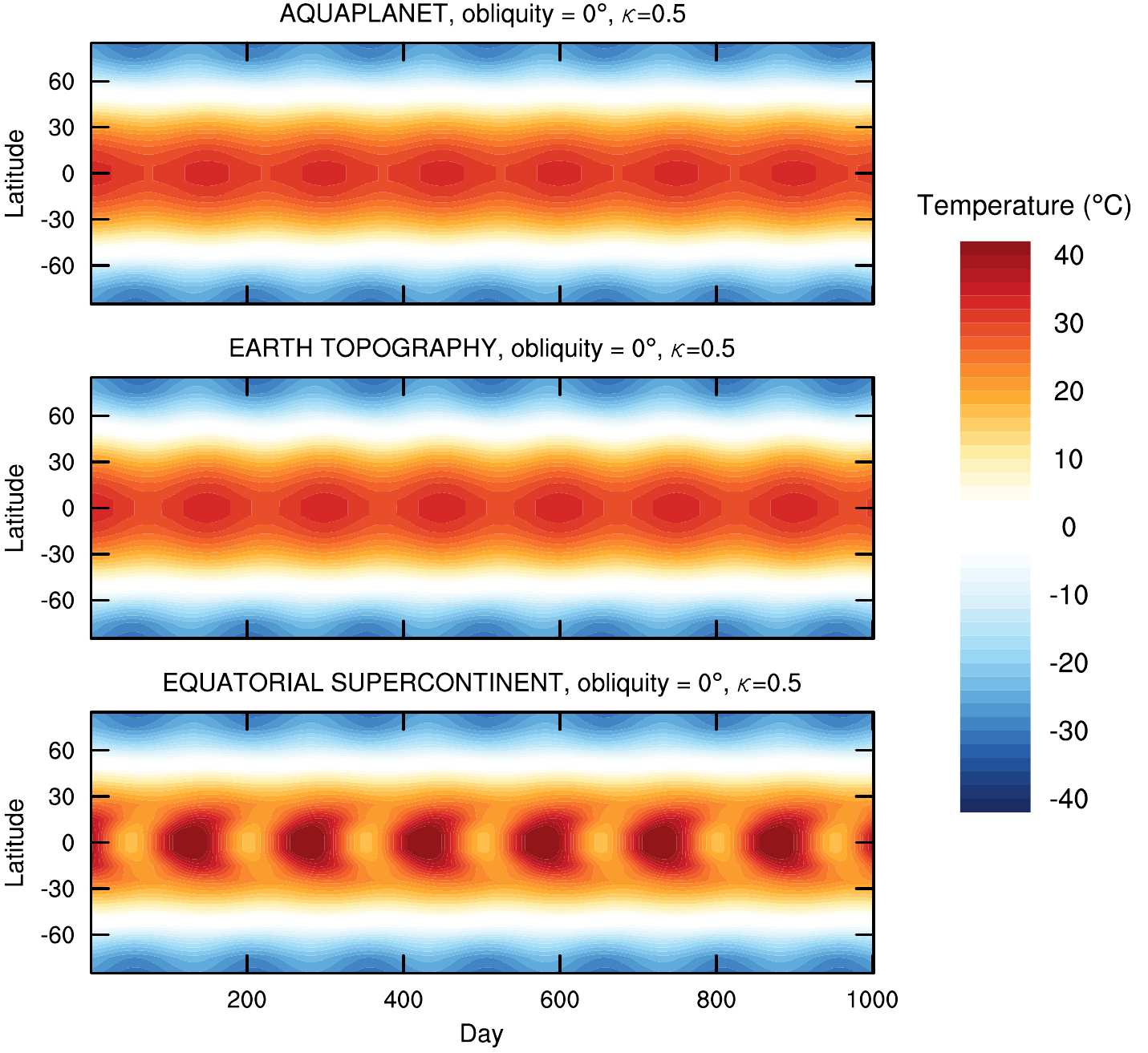}}
\caption{Steady state variations of the latitudinal distribution of planet temperature 
with time using the numerical EBM for a fictitious 
1\,M$_{\astrosun}$ primary and 0.5\,M$_{\astrosun}$ secondary case with $P_{bin}=150$ days.
All calculations assume strong circumbinary forcing ($\kappa=0.5$) and zero obliquity.
Periodicity in temperature is evident for global aquaplanet conditions with no topography (top),
Earth-like topography (middle), and an equatorial supercontinent as topography (bottom).
\label{fig:dailylattemp}}
\end{figure}

The choice of topography affects the amplitude and timing of temperature changes, as shown 
in Fig. \ref{fig:dailylattemp} for the circumbinary case with strong forcing ($\kappa = 0.5$)
and zero obliquity.
The top row shows the latitudinal distribution of temperature with time for an aquaplanet, comparable 
to the solid green line on Fig. \ref{fig:dailyavgtemp}. The bottom row likewise shows latitudinal 
temperature with time for a planet with equatorial supercontinent topography, comparable
to the dashed green line on Fig. \ref{fig:dailyavgtemp}. The middle row includes Earth topography,
which assumes $f_o = 0.7$ but distributes the land across the northern and southern hemispheres.
All cases show warm conditions along the equator and frozen conditions at latitudes
greater than about 50 degrees. The aquaplanet and Earth topography cases both include a substantial
fraction of ocean along the equator, which gives a higher heat capacity that allows equatorial
latitudes to remaining above freezing
over a complete circumbinary period. The equatorial supercontinent case 
has complete land cover and thus a much lower heat capacity along the equator,
which also remains above freezing at lower latitudes but shows a much
greater amplitude of oscillation in temperature along the equator.

\begin{figure}[ht!]
\centerline{\includegraphics[width=5.5in]{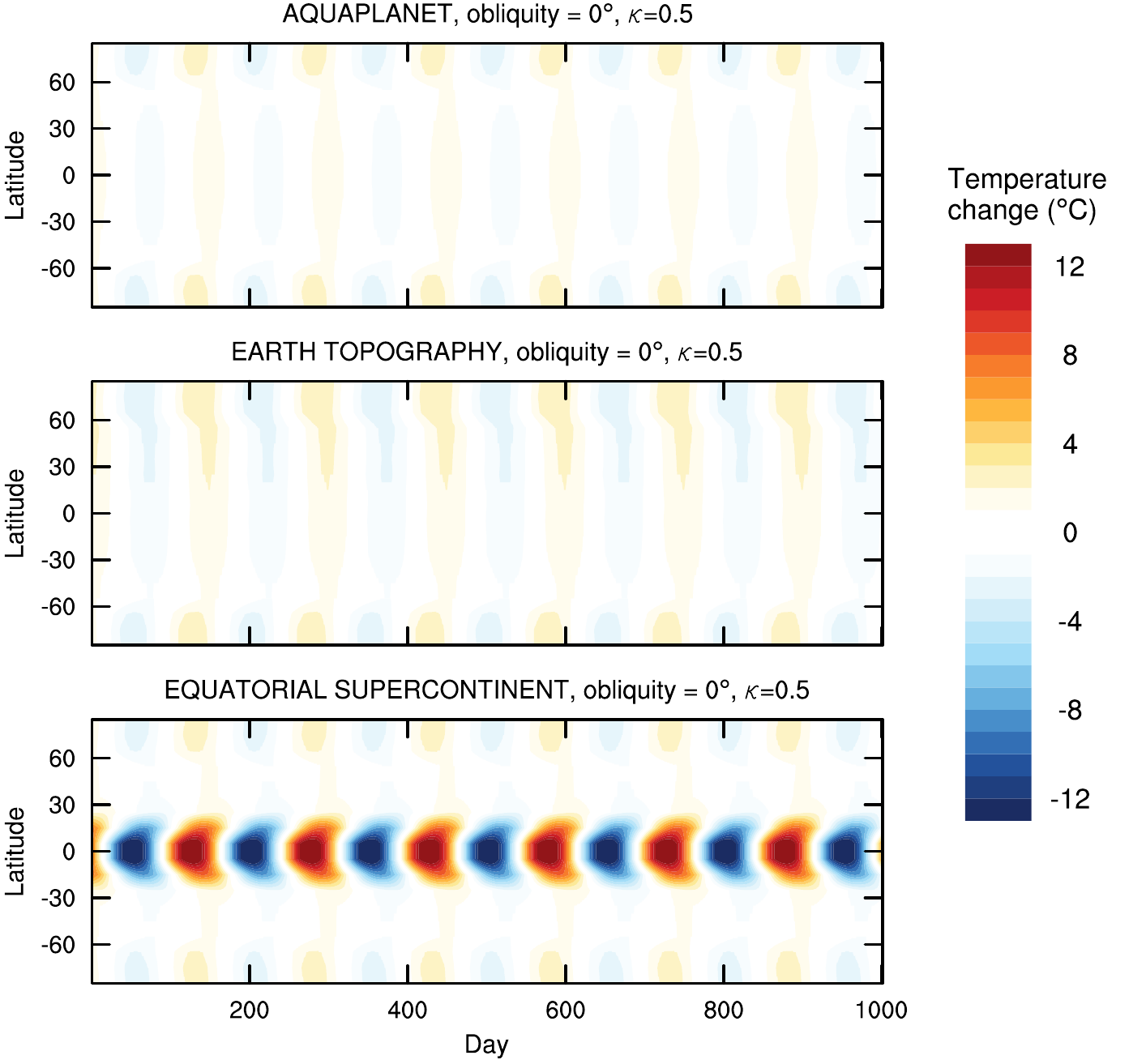}}
\caption{Steady state variations 
of the latitudinal distribution of planet temperature change 
with time using the numerical EBM for a fictitious 
1\,M$_{\astrosun}$ primary and 0.5\,M$_{\astrosun}$ secondary case with $P_{bin}=150$ days.
All calculations assume strong circumbinary forcing ($\kappa=0.5$) and zero obliquity.
Periodic changes in temperature are evident for global aquaplanet conditions with no topography (top),
Earth-like topography (middle), and an equatorial supercontinent as topography (bottom).
The temperature difference is calculated by taking the three cases from 
Fig. \ref{fig:dailylattemp} and subtracting the single-star temperature solutions
from the EBM with the same respective topography.
\label{fig:dailylattempdiff}}
\end{figure}

The transient warming and cooling induced by the circumbinary effect is illustrated in 
Fig. \ref{fig:dailylattempdiff}. This figure shows the temperature change obtained
by taking the three cases from Fig. \ref{fig:dailyavgtemp} and subtracting the corresponding
single-star temperature solutions of the numerical EBM. The difference shown in 
Fig. \ref{fig:dailylattempdiff} indicates symmetric temperature variation of a few degrees 
for the aquaplanet (top row), with the greatest variation occurring at 
polar latitude where ice caps form, and 
asymmetric temperature variation for Earth topography (middle row)
that reflects the larger concentration of land area in the northern hemisphere. 
The equatorial supercontinent topography (bottom row) shows changes of more than ten degrees
along the equator, with relatively little variation in the northern and southern oceans.
This is because the efficiency of the diffusive energy transport is enhanced over land compared to ocean 
(from Eqs. (\ref{eq:steadystatenumerical}) and (\ref{eq:heatcapacity})), which results in 
an increase in the amplitude of temperature variation on 
areas with large land fractions and a a decrease in variation over large oceans. 
The lack of a seasonal cycle also causes higher latitudes to experience a lower amplitude 
of stellar forcing, which contributes to a smaller variation in temperature.

\begin{figure}[ht!]
\centerline{\includegraphics[width=5.5in]{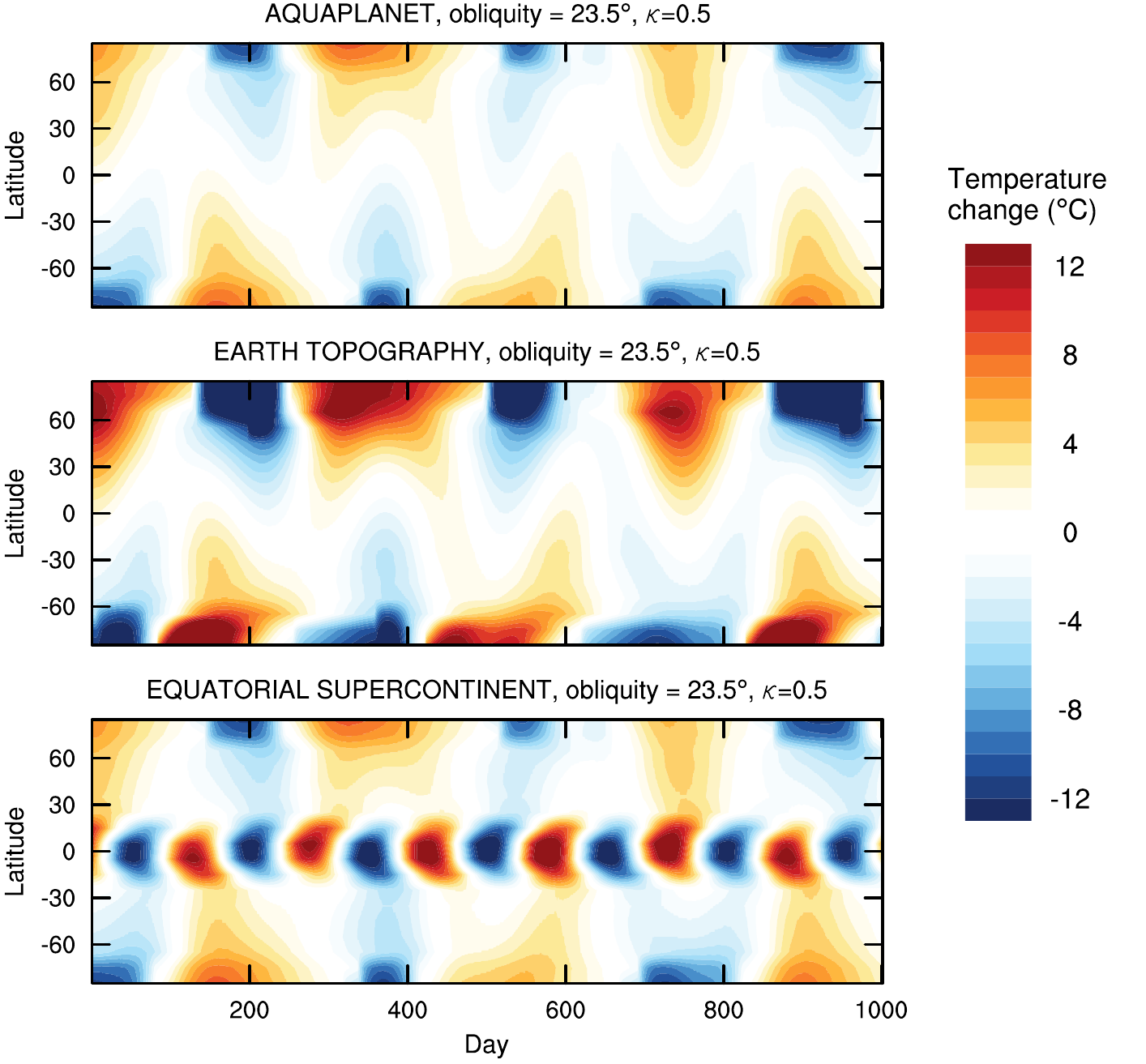}}
\caption{Steady state variations 
of the latitudinal distribution of planet temperature change 
with time using the numerical EBM for a fictitious 
1\,M$_{\astrosun}$ primary and 0.5\,M$_{\astrosun}$ secondary case with $P_{bin}=150$ days.
All calculations assume strong circumbinary forcing ($\kappa=0.5$) and 23.5$^\circ$ obliquity.
Periodic changes in temperature are evident for global aquaplanet conditions with no topography (top),
Earth-like topography (middle), and an equatorial supercontinent as topography (bottom).
The temperature difference is calculated by 
subtracting the single-star temperature solutions from the circumbinary solutions 
from the EBM with the same respective topography.
\label{fig:dailylattempdiff23obl}}
\end{figure}

This analysis has so far been restricted to a planet with zero obliquity, which
keeps the maximum extent of incident stellar radiation focused along the equator.
By contrast, a non-zero obliquity will result in a cycle that shifts
the maximum in circumbinary variation from the equator toward the poles 
with the seasons \citep{may2016examining}. We therefore consider a set of calculations
similar to those shown in Fig. \ref{fig:dailylattempdiff} but with Earth-like obliquity.
The temperature change obtained for a circumbinary planet with 23.5$^\circ$ obliquity
is shown in Fig. \ref{fig:dailylattempdiff23obl} for aquaplanet, Earth-like, and equatorial
supercontinent topographies and strong circumbinary forcing ($\kappa=0.5$). 
The aquaplanet case in Fig. \ref{fig:dailylattempdiff23obl} (top row) 
shows greater temperature change near the poles compared 
to the corresponding case in Fig. \ref{fig:dailylattempdiff}, due to the seasonal
extremes experienced at the planet's poles. Likewise, the Earth topography case (middle row) shows
amplified temperature change near the poles compared to the zero obliquity case, 
both of which are more extreme than the aquaplanet case due to the presence of continents
in both hemispheres. The equatorial supercontinent case in 
Fig. \ref{fig:dailylattempdiff23obl} (bottom row) shows the most similarity 
with the corresponding case in Fig. \ref{fig:dailylattempdiff}, although the 23.5$^\circ$ obliquity
case still shows seasonal variation of temperature near the poles. 
These results emphasize the strong control exerted by the ocean-to-land fraction
on the temperature variation that results on an Earth-like planet from the circumbinary 
gyration effect.

The analytic EBM indicates that the amplitude of temperature variation from the 
circumbinary gyration effect is sensitive to the effective heat capacity (Fig. \ref{fig:contour}),
but the numerical EBM illustrates that the latitudinal profile of heat capacity also
exerts a strong control on the maximum amplitude of warming 
(Figs. \ref{fig:dailylattempdiff} and \ref{fig:dailylattempdiff23obl}).
This occurs not only because
of the reduced heat capacity of land compared to ocean but also because diffusive 
meridional energy transport is enhanced over land. This feature may also be relevant
for planetary habitability, as large land areas (and thus any land-based life)
on circumbinary planets are be more likely
to experience temperature variations from the gyration effect than large oceans.

\subsection{Climate bistability}

\begin{figure}[t]
\centerline{\includegraphics[width=5.5in]{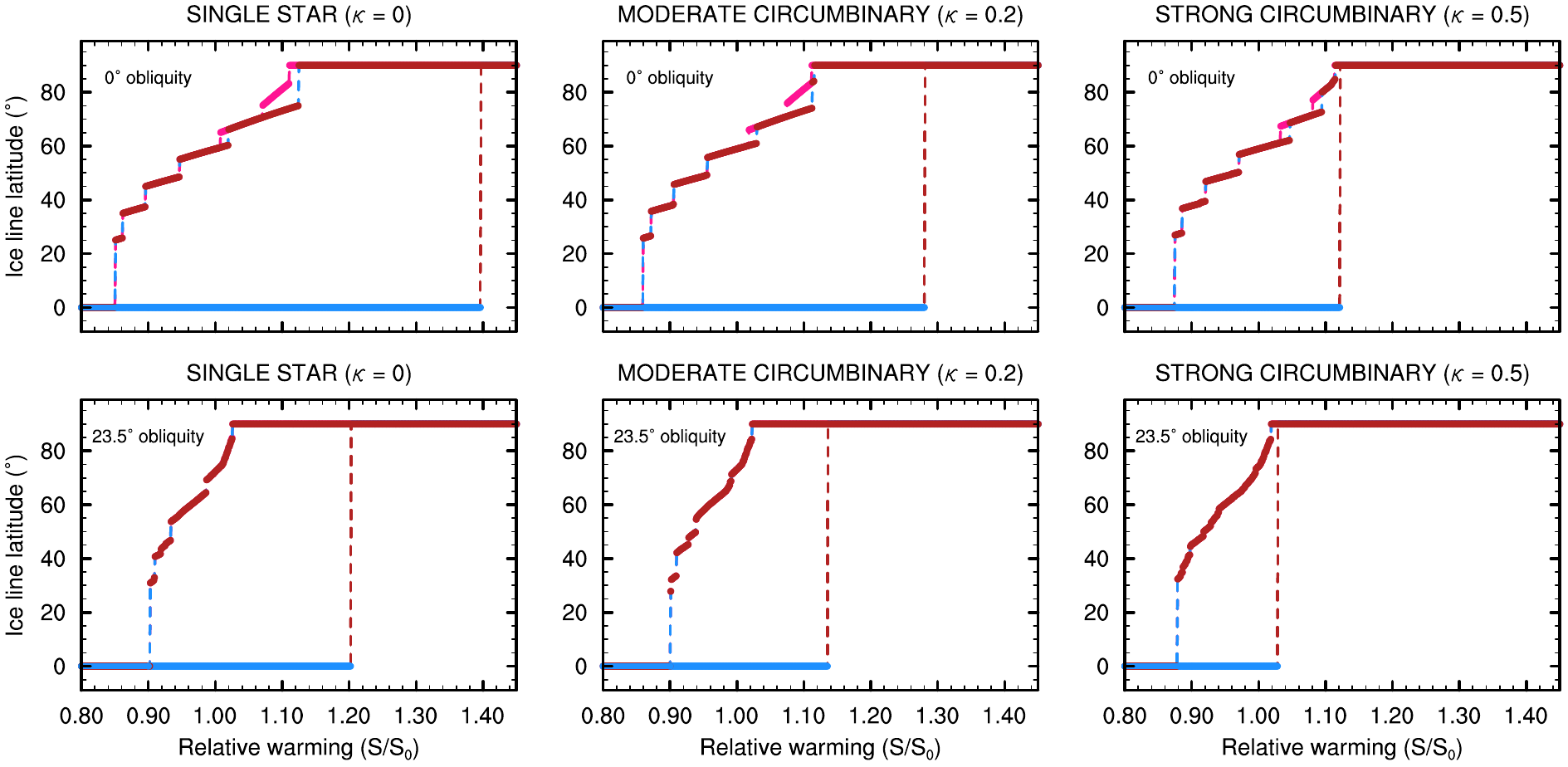}}
\caption{Hysteresis plots using the numerical EBM 
showing climate bistability for a planet with Earth-like topography
orbiting a single star ($\kappa=0$, left), a binary pair with moderate
forcing ($\kappa=0.2$, middle), and a binary pair with strong forcing ($\kappa=0.5$, right).
The top row shows cases with 0$^\circ$ obliquity and the bottom row shows cases
with 23.5$^\circ$ obliquity.
Climate states indicate solutions of the EBM when initialized with 
ice-covered (blue), ice-cap (red), and ice-free (pink) conditions. 
Dashed lines show discontinuous transitions between climate states. 
The width of the hysteresis loop between ice-covered and ice-free states 
narrows as the circumbinary gyration effect increases or as obliquity increases.
\label{fig:bistability}}
\end{figure}

The bistability of Earth-like climates 
is a well-known feature of EBMs \citep{north1981energy,caldeira1992susceptibility}
that also appears in many GCMs \citep{deconto2003rapid,ishiwatari2007dependence,voigt2010transition,wolf2017constraints}, with both warm and ice-covered
solutions available at a given value of stellar flux. 
The left panels of 
Fig. \ref{fig:bistability} illustrate this classic hysteresis loop for 0$^\circ$ 
obliquity (top) and 23.5$^\circ$ obliquity (bottom), with the blue
curve representing equilibrium EBM solutions initialized from ice-covered conditions
and the red curve indicating solutions initialized from ice-cap conditions. 
The dashed lines show transitions from an ice-covered to ice-free state (dashed red)
and from an ice-cap to ice-covered state (dashed blue). 
In general Fig. \ref{fig:bistability} describes a planet's climate state in terms of 
the relative warming and the previous climate state.
This hysteresis loop indicates
that large ice caps on Earth can remain stable until approximately 30 degrees, after which
the planet falls into global glaciation. 
This transition is not immediately reversible but 
instead requires a significant increase in relative warming (either a change 
in stellar flux or, equivalently in this case,
greenhouse gas forcing) before deglaciation can occur. On Earth, deglaciation from 
a global snowball condition seems to have occurred at least once, during the
Neoproterozoic Era \citep{hoffman1998neoproterozoic}.

The 0$^\circ$ obliquity case in Fig. \ref{fig:bistability}
also shows additional ice-free climate states in pink; 
this is known as the ``small ice cap'' solution, which describes the threshold at 
which a warming climate causes a small polar ice cap to become unstable and vanish.
This small ice cap instability has been observed in other EBMs 
\citep{north1984small,huang1992small} as well as some GCMs \citep{lee1995small,winton2006does}; 
however, the small ice cap is not as ubiquitous in climate
models as the more prominent large ice cap instability.
For the 0$^\circ$ obliquity case, the small ice cap instability occurs at approximately 
80 degrees, after which the planet transitions to an ice-free state. This transition
is also not immediately reversible; if relative
warming were to decrease on such a planet, then the climate would gradually cool and 
eventually fall into a large ice cap state.

Circumbinary planets show reduced bistability compared to their single star counterparts, as
shown in the middle and right columns of Fig. \ref{fig:bistability}. 
The cases with moderate forcing ($\kappa = 0.2$) show a reduction in the width of the 
hysteresis loop, which indicates a lower threshold for a frozen planet to deglaciate. 
This narrowing is even more pronounced in the strong forcing cases ($\kappa = 0.5$),
which features an abrupt transition from a glacial state to a warm, but nearly ice-cap,
state.
The temperature variation from the circumbinary gyration effect causes this 
narrowing of the hysteresis loop by reducing both the glaciation and deglaciation thresholds as 
the climate periodically exceeds mean temperature at all latitudes. 
The 0$^\circ$ obliquity cases show a wider hysteresis loop than the 23.5$^\circ$ obliquity cases,
as the latter includes seasonal forcing that cause additional melting at 
the summer pole \citep{gaidos2004seasonality,may2016examining}.
The increase in obliquity also causes the small ice cap instability to vanish in this model,
even though the moderate forcing ($\kappa = 0.2$) circumbinary case retains a small 
ice cap solution at zero obliquity. 

The zero obliquity cases may be less realistic for actual circumbinary planets, as the 
the orbital properties of the host stars and as the presence of other planets in the system
can place constraints on a planet's obliquity. In particular, spin-orbit coupling 
in such a system can cause a circumbinary planet to enter a Cassini State
that causes large variations in obliquity \citep{kostov2014kepler,forgan2015surface}.
Increasing the obliquity beyond the present-day Earth value considered in this study
would further increase the magnitude periodic temperature variation 
(Fig. \ref{fig:dailylattempdiff23obl}) and reduce the width of the 
hysteresis loop (Fig. \ref{fig:bistability}).
Earth-like circumbinary planets within the liquid water habitable zone may 
may therefore be less likely to be globally glaciated and more 
likely to exist in an ice-free or ice-cap state.

\section{Discussion and Conclusions}
\label{sec:disc}

We find that the habitability of circumbinary planets depends upon the ocean-to-land fraction
of the surface, as well as the amplitude of forcing from the gyration effect.
Terrestrial planets with larger land fractions than Earth, or a greater surface area 
of land at equatorial latitudes, should experience much greater shifts in temperature 
than planets with large ocean fractions. The lower effective heat capacity of land
as well as the enhanced efficiency of meridional energy transport make
such areas susceptible to greater changes in temperature from
variations in instellation. This behavior occurs because the atmospheres of planets we 
have considered are largely transparent to incident instellation (which is mostly at 
short wavelengths), with atmospheric greenhouse gases absorbing primarily at longer
infrared wavelengths. Such an assumption is generally valid for a wide range of plausible
terrestrial planets, even with atmospheres several times more dense than present-day Earth.
Circumbinary planets are expected to show reduced bistability, due to both the gyration
effect as well as high obliquity (Fig. \ref{fig:bistability}), which may suggest
that such planets may more easily deglaciate from a frozen snowball state. 

Our model calculations support the results of \citet{popp2017climate}, showing
small temperature variations from the gyration effect, 
when we examine a planet with global ocean coverage 
(Figs. \ref{fig:dailylattempdiff} and \ref{fig:dailylattempdiff23obl}, top row).
But when we include Earth-like land coverage
or an equatorial supercontinent 
(Figs. \ref{fig:dailylattempdiff} and \ref{fig:dailylattempdiff23obl},
middle and bottom rows, respectively), we find that the lower effective heat capacity
increases the extent of surface temperature variation. 
Such planets that experience a low to moderate variation in the circumbinary flux
are likely to experience changes
on a timescale analogous to a seasonal cycle,
which may not necessarily preclude habitability. 
Any Earth-like circumbinary planets that are eventually discovered may therefore be 
attractive candidates for spectroscopic characterization in the quest to discover potential
biosignatures with the next generation of space telescopes \citep{kiang2018exoplanet}. 

Circumbinary planets that lack large oceans are prone to 
experience a greater amplitude of temperature variation. 
The equatorial supercontinent cases shown in 
Figs. \ref{fig:dailylattemp}, \ref{fig:dailylattempdiff} and \ref{fig:dailylattempdiff23obl} 
exhibit periodic variation on a scale analogous to seasonal cycles but do not freeze
along the equator, so habitability cannot be precluded on such planets. 
However, the equatorial supercontinent case is not a pure land planet and  
includes ocean at the poles, with $f_o = 0.7$. The numerical EBM
became numerically unstable at lower values of lower values of $f_o$, which indicates
the high sensitivity of this model to the planet's effective heat capacity. 
Planets with much larger land fractions could experience even greater
variation in temperature that could lead to freezing conditions along the equator.
While such changes could pose challenges for biology,
the freeze-thaw cycle on such planets could instead provide a selection pressure
that helps to drive biological evolution. The magnitude of temperature change on such 
planets will help predict whether such behavior is a detriment to habitability and will 
require more detailed calculations using a GCM tailored to the particular system.

M-dwarf stars emit a substantial fraction of radiation at infrared wavelengths, so 
a planet orbiting a circumbinary pair of M-dwarf stars would be able to absorb a greater
fraction of incident instellation with its atmosphere. A planet with a sufficiently dense 
atmosphere around such a pair of low-mass stars might even absorb enough incoming radiation
that periodic variations on the surface are damped and overall less sensitive to the ocean
fraction. This would also drive complex behavior, such as changes in the substellar cloud
deck, that cause further feedbacks on climate. 
Testing this scenario will require calculations with a GCM that include
stellar spectra from a low-mass circumbinary pair.

This study has explored the extent of periodic temperature variation on Earth-like
circumbinary planets; however, diurnal variations in climate cannot be captured with either
the analytic or numerical EBM. 
Similarly, the representation of topography in the numerical EBM is based only on the 
ocean-to-land fraction at each latitude band, with no regard to the distribution 
of land with longitude. 
Previous GCM studies of circumbinary planets have assumed uniform surface
conditions, and focus on spatially averaged quantities,  
but localized warming effects over the diurnal cycle and across continents could limit
habitability in certain regions on the surface; for example, the surface habitability
limits suggested by \citet{sherwood2010adaptability} are based upon the ability for 
humans and other mammals to efficiently dissipate metabolic heat.
Further work with GCMs will therefore provide quantitative constraints on the 
expected climates of terrestrial circumbinary planets that experience extreme forcing from
their binary star host.

\section*{Appendix: Analytic Energy Balance Model}
This appendix provides the analytic solution of the energy balance model equation
with sinusoidal forcing in Eq. (\ref{eq:EBM2}).
We begin by writing the homogeneous part of Eq. (\ref{eq:EBM2}), 
\begin{equation}
\frac{dT}{dt}+\frac{B}{C}T=0.\label{eq:EBM-homo}
\end{equation}
The solution to Eq. (\ref{eq:EBM-homo}) gives $T_{g}\left(t\right)=X\exp\left(-Bt/C\right)$,
where $X$ is a constant. This is the general solution to Eq. (\ref{eq:EBM2}).
We next find the particular solution of Eq. (\ref{eq:EBM2}) by assuming
a solution in the form $T_{p}\left(t\right)=Y\sin\left(\omega t\right)+Z\cos\left(\omega t\right)+W$,
where $Y$, $Z$, and $W$ are constants. Substituting this expression
for $T_{p}\left(t\right)$ into Eq. (\ref{eq:EBM2}) gives 
\begin{equation}
\left[BY-\omega{C}{Z}\right]\sin\left(\omega t\right)+\left[BZ+\omega{C}{Y}\right]\cos\left(\omega t\right)=S_{0}\left(1-\alpha\right)\left[1+\kappa\sin\left(\omega t\right)\right]-A-BW.\label{eq:particular}
\end{equation}
Setting the constant terms of Eq. (\ref{eq:particular}) equal to
each other gives the value of $W$, as
\begin{equation}
W=\frac{S_{0}\left(1-\alpha\right)-A}{B}.\label{eq:Wval}
\end{equation}
The constants $Y$ and $Z$ can be found by equating the coefficients
of the trigonometric functions on both sides of Eq. (\ref{eq:particular}), which gives the values
\begin{equation}
Y=\frac{S_{0}\left(1-\alpha\right)\kappa B}{\omega^{2}C^{2}+B^{2}},\label{eq:Yval}
\end{equation}
and
\begin{equation}
Z=\frac{-S_{0}\left(1-\alpha\right)\kappa\omega C}{\omega^{2}C^{2}+B^{2}}.\label{eq:Zval}
\end{equation}
The full solution to Eq. (\ref{eq:EBM2}) is given as the sum of the
general and particular solutions, $T\left(t\right)=T_{g}\left(t\right)+T_{p}\left(t\right)$.
We can now write this solution as 
\begin{equation}
T\left(t\right)=X\exp\left(\frac{-Bt}{C}\right)+Y\sin\left(\omega t\right)+Z\cos\left(\omega t\right)+W.\label{eq:fullgeneral}
\end{equation}
The values of $Y$, $Z$, and $W$ are known. To find the value of
$X$, we assume $T\left(t\right)=T_{0}$ at $t=0$. This boundary
condition reduces Eq. (\ref{eq:fullgeneral}) to $T_{0}=X+Z+W$, which
gives 
\begin{equation}
X=T_{0}+S_{0}\left(1-\alpha\right)\left[\frac{\kappa\omega C}{\omega^{2}C^{2}+B^{2}}-\frac{1}{B}\right]+\frac{A}{B}.\label{eq:Xval}
\end{equation}
This solution can be verified by substituting Eq. (\ref{eq:fullgeneral})
into Eq. (\ref{eq:EBM2}) using the respective values of $W$, $X$,
$Y$, and $Z$ from Eqs. (\ref{eq:Wval}), (\ref{eq:Xval}), (\ref{eq:Yval}),
and (\ref{eq:Zval}). 

We can simplify Eq. (\ref{eq:fullgeneral}) by examining the steady state solution 
as $t$ becomes large and the general solution to Eq. (\ref{eq:EBM2}) vanishes, $T_{g}(t) \approx 0$.
The particular solution to Eq. (\ref{eq:EBM2}), $T_{p}(t)$, can be reduced by noting that
$\omega{C}\gg{B}$ for the range of paramters considered in this study. By first rewriting the
particular solution as $T_{p} = \sqrt{Y^{2}+Z^{2}}\cos\left(\omega t-\arctan\tfrac{Z}{Y}\right) + W$, 
we arrive at an expression for the steady state solution to Eq. (\ref{eq:EBM2}):
\begin{equation}
T\left(t\right)\approx\frac{S_{0}\left(1-\alpha\right)\kappa}{\omega C}\cos\left(\omega t\right)+T_{0}.
\end{equation}

%

\acknowledgments
The data used for this paper can be accessed at https://doi.org/10.6084/m9.figshare.10279790.
The authors gratefully acknowledge funding from the NASA
Habitable Worlds program under award 80NSSC17K0741. 
J.H., E.T.W., and R.K.K. also acknowledge funding from the Virtual Planetary Laboratory 
under NASA award 80NSSC18K0829, and R.K.K. and V.K. acknowledge funding 
from the Sellers Exoplanet Environments Collaboration.
W.F.W. gratefully thanks John Hood Jr. for his support of exoplanet research at SDSU.
Any opinions, findings, and conclusions or recommendations expressed in this material
are those of the authors and do not necessarily reflect the views of
NASA.

%
%
%
%
%
%
%
%
%

\bibliography{main.bib}

%

\end{document}